\documentclass{emulateapj}
\usepackage{apjfonts}
\slugcomment{Accepted by  \textsc{the Astrophysical Journal}}

\begin{document}

\title{Generation of Type I X-ray Burst Oscillations by Unstable Surface Modes}
\shorttitle{Generation of Burst Oscillations}

\author{Ramesh Narayan and Randall L.\ Cooper\altaffilmark{1,2}}
\shortauthors{Narayan \& Cooper}
\affil{Harvard-Smithsonian Center for Astrophysics, 60 Garden 
Street, Cambridge, MA 02138}
\altaffiltext{1}{Kavli Institute for Theoretical Physics, Kohn Hall, University of 
California, Santa Barbara, CA 93106}
\altaffiltext{2}{KITP Graduate Fellow}

\email{rnarayan@cfa.harvard.edu, rcooper@cfa.harvard.edu}

\begin{abstract}

The {\it Rossi X-ray Timing Explorer} has detected nearly coherent
oscillations in the tails of type I X-ray bursts from 17 low-mass
X-ray binaries.  The oscillations are thought to be generated by
brightness fluctuations associated with a surface mode on the rotating
neutron star.  The mechanism that drives the modes is, however, not
understood, since the burning layer is stable to thermal
perturbations.  We show here via a linear perturbation analysis that,
even under conditions when pure thermal perturbations are stable,
nonradial surface modes may still be unstable by the $\epsilon$
mechanism.  Specifically, we find that, if helium-burning reactions
supply a reasonable fraction of the outgoing flux during burst decay,
nonradial surface modes will grow in time.  On the other hand, the
same modes are likely to be stable in the presence of hydrogen burning
via the $rp$-process.  The results naturally explain why oscillations
in the decay phase of type I X-ray bursts are detected only from
short-duration bursts.

\end{abstract} 


\section{Introduction}\label{introduction}

Type I X-ray bursts are thermonuclear explosions on the surfaces of
accreting neutron stars, triggered by unstable hydrogen or helium
burning \citep{Betal75,GH75,Getal76,BCE76,WT76,J77,MC77,LL77,LL78}.
They typically have fast rise times of $\sim 1$ s, exponential-like
decays lasting $\sim 10$--$100$ s, and recurrence times of a few hours
to days \citep[for recent reviews, see][]{C04,SB06}.  

\citet{Setal96} were the first to detect coherent oscillations during
a type I X-ray burst.  Since their initial discovery, oscillations
have been seen in lightcurves from helium-triggered bursts in 17
sources \citep[Strohmayer \& Bildsten 2006 and references
therein;][]{BSMS06,Ketal07,B07}, as well as from one carbon-triggered
superburst in 4U 1636--536 \citep{SM02}.  Oscillations occur during
the burst rise as well as the burst tail.  The oscillation frequencies
often increase by a few percent during the course of a burst,
asymptoting to a value that is found to be stable over years to within
a few parts in $10^{3}$ \citep[e.g.,][]{GHSC02,MCGP02}.  The
asymptotic frequency is thought to correspond to the neutron star spin
frequency \citep[e.g.,][]{SM99}.  As of this writing, burst
oscillations have been detected only from sources with large values of
the inferred accretion rate \citep{MFMB00,MGC04,vanSetal01,F01}.

Oscillations during the burst rise are believed to be caused by the
rotational modulation of a growing hot spot \citep{SZS97,SZSWL98}.
The thermonuclear instability that causes a burst is almost certainly
triggered initially at a point \citep{J78,S82}, and the resulting hot
spot presumably then expands and engulfs the entire surface in $\sim
1$ s, the burst rise time \citep{FW82,B95,Zetal01,SLU02}.  At
relatively low accretion rates, ignition most likely occurs at the
equator, and the resulting thermonuclear flame quickly spreads in
longitude and generates an axisymmetric belt around the neutron star.
Oscillations are not expected to be detected because there is no
azimuthal asymmetry \citep{SLU02}.  At high accretion rates, however,
ignition most likely occurs off the equator \citep{CN07}.  The
resulting thermonuclear flame propagates in longitude much more slowly
\citep{SLU02}, creating a relatively long-lived non-axisymmetric hot
spot and leading to observable oscillations.

While this simple model explains the oscillations observed during
burst rise, it does not work for oscillations during burst decay since
the data suggest that the entire neutron star surface radiates during
this time.  The most promising explanation for oscillations in the
burst tail is that they are generated by excited surface modes on the
neutron star \citep{MT87,L04,H04,H05,LS05,PB06}.  In an important
paper, \citet{H04} identified surface $r$-modes as the most promising
candidate.  These modes travel backward (east to west) in the
corotating frame and so the observed oscillation frequency is reduced
relative to the neutron star spin frequency.  As the neutron star
surface cools during the burst decay, the mode frequency decreases,
thus explaining why the observed oscillation frequency increases.
However, the expected frequency drift that \citet{H04} calculated is
significantly larger than the observed drifts.  \citet{PB05b} showed
that a surface mode in the burning layer transitions into a crustal
interface mode \citep{PB05a} during the burst decay.  This reduces the
expected frequency drift to values that are in accord with the
observed drifts.

There is, however, one major unsolved problem: It is not clear what
excites the surface mode during the burst decay and what keeps it
active several seconds after the burst was initially triggered.  It is
highly unlikely that oscillations in the burst tail are produced by
remnant asymmetry left over from the initial aspherical ignition.  The
cooling time at the burst peak is much less than the burst duration,
so any initial asymmetry would have dissipated by this time.  Also,
since the modes identified by Heyl occupy only a small area of the
neutron star surface around the equator, one requires huge oscillation
amplitudes in order to explain the $\sim10\%$ flux variations
observed.  It is hard to see how such a large amplitude could be
maintained late in the burst if the modes were excited only at the
time of ignition.

In our view, a more reasonable explanation of the observations is that
the mode responsible for oscillations during burst decay is {\it
unstable}.  An instability would naturally explain the large
oscillation amplitude.  It would also explain why some type I X-ray
bursts exhibit oscillations late in the burst decay, well after the
time of the burst peak, when the thermonuclear flame has engulfed the
entire stellar surface.

In this investigation, we analyze the surface modes of a bursting
neutron star, paying particular attention to the heating and cooling
processes in the burning layer.  We identify the conditions under
which modes will grow and show that these conditions are likely to be
satisfied for at least some type I X-ray bursts.  Our results
naturally explain why oscillations in the tails of bursts from
nonmagnetic accreting neutron stars are excited specifically in short
bursts.

We begin in \S \ref{modedynamics} by deriving the equations that
govern the mode dynamics.  In \S \ref{thermalstructure} we discuss the
thermal structure of the accreted layer during the burst peak using a
simple one-zone model.  We combine the mode equations and the thermal
structure model in \S \ref{modeanalysis} and analyze the resulting
modes.  We then determine the stability of these modes in \S
\ref{growthrate}.  We derive the criterion for instability and
demonstrate that surface modes are likely to be excited during the
decay phase of type I X-ray bursts.  We conclude in \S
\ref{discussion} with a discussion of the results.  It should be noted
that the mechanism for mode instability described here is different
from the shear instability of zonal flows discussed by \citet{C05b}.
It is similar in spirit to the model of \citet{PB04}, who analyzed the 
stability of nonradial oscillations during thermally stable nuclear 
burning on the surface of a neutron star accreting at a
super-Eddington rate, and to the models of \citet{MT87} and \citet{SL96}, 
who found that the $\epsilon$ mechanism could drive surface modes 
during thermally unstable nuclear burning.

\section{Height-Integrated Approximation and Mode Dynamics}
\label{modedynamics}

We make two principal simplifications in our analysis of the
oscillation modes of the burning surface layer of a rapidly rotating
accreting neutron star:

\noindent
(I) We assume that the mode is concentrated within a narrow equatorial
belt.  This allows us to ignore the spherical geometry of the stellar
surface and thereby obtain analytic solutions.  As shown in the
analysis of Longuet-Higgins (1968, see also Bildsten, Ushomirsky \&
Cutler 1996), the modes are indeed equatorially concentrated whenever
the angular frequency of the star $\Omega$ is much larger than the
characteristic mode frequency.  This condition is satisfied for
rapidly rotating neutron stars with $\Omega \sim {\rm few}\,\times
10^3$ radians per second.

\noindent
(II) We assume that each column of matter is always in radial
hydrostatic equilibrium.  This is a safe assumption since the vertical
sound crossing time of the layer ($\ll 1 \,\mu$s) is much shorter than
the typical mode period ($\sim$ tens of ms).

Making use of assumption I, we treat the equatorial belt of the star
as a cylindrical surface of radius $R$ and surface gravity $g$.  We
use coordinates $x, ~y, ~z$ to represent linear displacements in (i)
the azimuthal direction (east-west), (ii) the meridional direction
(north-south) with $y=0$ located at the equator, and (iii) the radial
direction (vertical) with $z=0$ located at the bottom of the burning
layer, respectively.  The coordinate $x$ is periodic with a period of
$2\pi R$; by assumption I, the range of $y$ of interest to us is small
compared to $R$; since the layer is generally very thin ($\sim 100$
cm), the range of $z$ is even smaller.  Thus, over the volume of
interest, the parameters $R$ and $g$ may be assumed constant.
However, the Coriolis force, which depends on the projection of the
stellar angular velocity vector on the local normal to the surface,
does vary with position.  Making use of assumption I, we write the
Coriolis parameter $f$ as
\begin{equation}
f = 2\Omega {y\over R},
\label{f}
\end{equation}
which is valid for small excursions around the equator.  This is the
well-known $\beta$-plane approximation used in the study of Rossby
waves in the atmosphere \citep[e.g.,][]{H02}.

In equilibrium, the density, temperature, and pressure of the layer
are independent of $x$, $y$, and time $t$ and are described by the
functions $\rho_0(z)$, $T_0(z)$, $p_0(z)$, respectively.  Let us
represent the first-order perturbations of the density, temperature
and pressure by $\rho'$, $T^{'}$, and $p'$, each of which is a
function of $x, ~y, ~z$, and $t$.  The equilibrium configuration has
no motions as viewed in the rotating frame.  The perturbations do
involve motions, and we write the three components of the velocity
perturbation as $u$, $v$, $w$.  The continuity equation and the three
components of the momentum equation then take the form
\begin{eqnarray}
{\partial\rho'\over\partial t} + \rho_0\left({\partial u\over\partial x}
+{\partial v\over\partial y}+{\partial w\over\partial z}\right) &=& 0, \label{3d1}\\
\rho_0{\partial u\over\partial t} + {\partial p'\over\partial x}
-\rho_0 fv &=& 0, \\
\rho_0{\partial v\over\partial t} + {\partial p'\over\partial y}
+\rho_0 fu &=& 0, \\
\rho_0{\partial w\over\partial t} + {\partial p'\over\partial z} 
+g\rho' &=& 0. \label{3d4}
\end{eqnarray}

We now make use of assumption II and ignore the dynamics in the
$z$-direction.  That is, we assume that the pressure $p$ at
any given depth $z$ in a local patch of the plasma is related simply to
the local column density $\Sigma$ above this $z$ by
\begin{equation}
p(x,y,z,t) = g\int_z^{h}\rho(x,y,\tilde{z},t)d\tilde{z},
\end{equation}
where $z=h$ corresponds to the surface of the layer.  Let us define
$\Sigma_0$ to be the equilibrium column density of the burning layer
(a constant), $\Sigma^{'}(x,y,t)$ to be the perturbed column density,
and $\Sigma$ to be the sum of these two.  Integrating over $z$,
\begin{eqnarray}
\Sigma &=& \int_0^h \rho dz = \Sigma_0 + \Sigma^{'}, \\
h &=& h_0 + h',
\end{eqnarray}
where, as before, the subscript 0 corresponds to the equilibrium value
of a quantity and a prime corresponds to the linear perturbation.
Similarly, we define a height-integrated pressure
\begin{equation}
P = \int_0^h pdz = P_0 + P^{'}.
\end{equation}

We now integrate equations (\ref{3d1})--(\ref{3d4}) over $z$ to obtain
the following height-integrated dynamical equations:
\begin{eqnarray}
{\partial\Sigma^{'}\over\partial t} + \Sigma_0\left({\partial u\over\partial x}
+{\partial v\over\partial y}\right) &=& 0, \label{2d1}\\
\Sigma_0{\partial u\over\partial t} + {\partial P^{'}\over\partial x}
-2\Omega\Sigma_0 {y\over R}v &=& 0, \label{2d2}\\
\Sigma_0{\partial v\over\partial t} + {\partial P^{'}\over\partial y}
+2\Omega\Sigma_0 {y\over R}u &=& 0, \label{2d3}
\end{eqnarray}
where we have used equation (\ref{f}) for $f$.  In carrying out the
height-integration, we have assumed that $u$ and $v$ are independent
of $z$.  This is a reasonable assumption so long as the mode under
consideration has no nodes in the vertical direction.  Henceforth, our
work is restricted exclusively to such modes (which are likely to be
the most interesting ones for burst oscillations).

We decompose the linear perturbations into modes in the usual way and
write
\begin{equation}
\Sigma^{'}(x,y,t)=\Sigma^{'}(y) \exp(-i\omega t+imx/R),
\end{equation}
with similar expressions for $P^{'}$, $T^{'}$, $u$, $v$, etc.  Here,
$\omega$ is the mode frequency (as seen in the rotating frame), which
is in general complex, and $m$ is the azimuthal wave number, an
integer.  Substituting in equations (\ref{2d1})--(\ref{2d3}), we
obtain the following mode equations:
\begin{eqnarray}
-{i\omega\over \Sigma_0}\Sigma^{'} + {im\over R} u
+{dv\over dy} &=& 0, \label{dyn1}\\
-i\omega u + {im\over R\Sigma_0}P^{'}
-{2\Omega\over R} yv &=& 0, \label{dyn2}\\
-i\omega v + {1\over\Sigma_0}{d P^{'}\over d y}
+{2\Omega\over R} yu &=& 0. \label{dyn3}
\end{eqnarray}
To close this system of equations we need an equation of state
relating the effective two-dimensional height-integrated 
pressure perturbation $P^{'}$
and the column density perturbation $\Sigma^{'}$.  If we assume a
simple adiabatic or polytropic equation of state, all the mode
frequencies will be real and there will be no instability.
Instability is possible only as a result of an interaction between the
mode dynamics and thermal effects associated with nuclear reactions in
the layer.  This motivates the discussion in the next section.

\section{Thermal Structure of the Burning Layer}\label{thermalstructure}

For simplicity, we make use of a one-zone approximation for the
vertical structure of the accreted layer \citep[e.g.][]{FHM81,B98}.  
Thus, at each $(x,y)$, we assume
that the matter has constant density and write the height of the layer
simply as
\begin{equation}
h_0 = {\Sigma_0\over\rho_0}, \quad
h = {\Sigma\over\rho}, \quad
h' \equiv h-h_0 = {\Sigma_0\over\rho_0} \left({\Sigma^{'}\over\Sigma_0}
-{\rho'\over\rho_0}\right).
\label{h}
\end{equation}
We assume an ideal gas equation of state with adiabatic index
$\gamma$,
\begin{equation}
p={\rho kT\over\mu m_p}, \quad s=c_V\ln\left(T\over\rho^{\gamma-1}
\right), \quad c_V = {k\over(\gamma-1)\mu m_p},
\label{eos}
\end{equation}
where $s$ is the entropy per unit mass and $c_V$ is the specific heat
per unit mass at constant volume.  
Since $p=g\Sigma$ at the bottom of the
layer (by the assumption of vertical hydrostatic equilibrium), we
obtain by perturbing the equation of state that
\begin{equation}
{T^{'}\over T_0} = {\Sigma^{'}\over\Sigma_0} - {\rho'\over \rho_0}.
\label{Teq}
\end{equation}

Combining the one-zone approximation with the condition for
hydrostatic equilibrium, we find for the layer height
\begin{equation}
h_0={kT_0\over g\mu m_p}, \quad h={kT\over g\mu m_p},
\end{equation}
and for the height-integrated pressure
\begin{equation}
P=\int_0^hpdz=\int_0^h g\rho(h-z)dz = {1\over2}g \Sigma h.
\end{equation}
Differentiating the latter, we obtain
\begin{equation}
P^{'} = {1\over2}g \Sigma_0 h_0 \left({\Sigma^{'}\over\Sigma_0}
+{h'\over h_0}\right) = {g\Sigma_0^2\over2\rho_0}\left(
2{\Sigma^{'}\over\Sigma_0}-{\rho'\over\rho_0}\right).
\label{pressurepert}
\end{equation}
To convert this to an effective equation of state of the form $P^{'} =
P^{'}(\Sigma^{'})$, we need a relation between $\rho'$ and
$\Sigma^{'}$.  The rest of the section is devoted to deriving this
relation.

The entropy per unit volume of the matter in the layer evolves according
to
\begin{equation}
\rho T{\partial s\over\partial t} = \rho\epsilon 
+{\partial F\over \partial z},
\end{equation}
where $\epsilon$ is the energy generation rate per unit mass through
nuclear reactions, and $F$ is the energy flux.  For the latter we use
the radiative diffusion equation
\begin{equation}
F = {ac\over3\kappa}{\partial T^4\over\partial\Sigma},
\end{equation}
where $\kappa$ is the opacity of the plasma.  Using the one-zone
approximation and integrating over $z$, these relations become
\begin{equation}
\Sigma T{\partial s\over\partial t} = \Sigma\epsilon -(F- F_b),
\qquad F = {acT^4\over3\kappa\Sigma}.
\label{entropy}
\end{equation}
Here, $F_b$ is the flux flowing into the base of the layer from the
interior of the star, which we consider to be a constant, and $F$ is
the flux escaping from the upper surface of the layer.

We now consider linear perturbations of the entropy equation.  Let us
begin with the entropy term.  For small perturbations
\begin{equation}
s' = c_V\left[{T^{'}\over T_0} - {(\gamma-1)\rho'\over\rho_0}\right].
\label{entropy2}
\end{equation}
Substituting for $T^{'}/T_0$ from equation (\ref{Teq}) and setting
$\partial/\partial t=-i\omega$, the perturbation of the entropy term
in equation (\ref{entropy}) becomes
\begin{equation}
\left(\Sigma T{\partial s\over\partial t}\right)^{'} = -i\omega
gh_0\Sigma_0\left[{\Sigma^{'}\over(\gamma-1)\Sigma_0}
-{\gamma\rho'\over(\gamma-1)\rho_0}\right].
\label{entropypert}
\end{equation}

The nuclear energy generation rate $\epsilon$ is in general a function
of both density and temperature.  For small perturbations around the
equilibrium, let us write this dependence via two indices $\eta$ and
$\nu$,
\begin{equation}
\epsilon(\rho,T) =\epsilon_0\left({\rho\over\rho_0}\right)^\eta
\left({T\over T_0}\right)^\nu
\equiv gh_0\Omega_h\left({\rho\over\rho_0}\right)^\eta \left({T\over
T_0}\right)^\nu,
\end{equation}
where for later convenience we have defined $\epsilon_0\equiv
gh_0\Omega_h$ such that $\Omega_h^{-1}$ is the characteristic heating
time of the matter through nuclear reactions.  The perturbation of the
heating term in the entropy equation (\ref{entropy}) is then
\begin{equation}
(\Sigma\epsilon)^{'} = \Sigma_0 gh_0\Omega_h\left[
(\nu+1){\Sigma^{'}\over\Sigma_0}+(\eta-\nu){\rho'
\over\rho_0}\right]. \label{heatpert}
\end{equation}

Similarly, for the radiative cooling term $F$, let us write the
dependence of the opacity $\kappa$ on density and temperature as
\begin{equation}
\kappa(\rho,T) = \kappa_0\left({\rho\over\rho_0}\right)^\xi
\left({T\over T_0}\right)^\zeta,
\end{equation}
and also define $\Omega_c$ in terms of the equilibrium escaping flux
$F_0$,
\begin{equation}
F_0 \equiv \Sigma_0 gh_0 \Omega_c,
\end{equation}
so that $\Omega_c^{-1}$ is the characteristic cooling time of the
layer.  The perturbation of the cooling term then gives
\begin{equation}
F^{'} = \Sigma_0 gh_0\Omega_c\left[
(3-\zeta){\Sigma^{'}\over\Sigma_0}-(4+\xi-\zeta){\rho'
\over\rho_0}\right]. \label{coolpert}
\end{equation}

Substituting equations (\ref{entropypert}), (\ref{heatpert}) and
(\ref{coolpert}) in the perturbed entropy equation
\begin{equation}
\left(\Sigma T{\partial s\over\partial t}\right)^{'} = 
(\Sigma\epsilon)^{'} - F^{'},
\end{equation}
and rearranging terms, we obtain an expression for $\rho'/\rho_0$ in
terms of $\Sigma^{'}/\Sigma_0$.  Substituting this in equation
(\ref{pressurepert}) we finally obtain the effective equation of state
of the matter in the layer.  The result is
\begin{equation}
P^{'} = Agh_0 \Sigma^{'},
\label{effeos}
\end{equation}
where $A$ is the following dimensionless quantity,
\begin{eqnarray} &&
A = {(2\gamma-1)\over2\gamma} 
\times  \nonumber \\ \nonumber \\ &&
\left[{1+i[(\gamma-1)/(2\gamma-1)][(\Omega_h/\omega)(2\eta-\nu+1)
+(\Omega_c/\omega)(5+2\xi-\zeta)] \over
1+i[(\gamma-1)/\gamma][(\Omega_h/\omega)(\eta-\nu)
+(\Omega_c/\omega)(4+\xi-\zeta)]}
\right]. 
\nonumber \\ \nonumber \\
\label{A}
\end{eqnarray}
For a given mode, $A$ is a constant; it is, in general, complex.

For pure adiabatic perturbations (i.e., no heating or cooling), $A$
simplifies to $(2\gamma-1)/2\gamma$ and the height-integrated pressure
perturbation satisfies
\begin{equation}
{P^{'}\over P_0} = {(2\gamma-1)\over\gamma} {\Sigma^{'}\over\Sigma_0}.
\end{equation}
Thus, the two-dimensional fluid behaves as if it has an effective adiabatic
index
\begin{equation}
\gamma_{\rm eff} = {(2\gamma-1)\over \gamma}.
\end{equation}
If the fluid is incompressible in three dimensions
($\gamma\to\infty$), for example, the height-integrated
two-dimensional system has an effective index $\gamma_{\rm eff} =2$.
By noting this correspondence, one can easily map many of the
equations in the present paper to corresponding results in
Longuet-Higgins's (1968) analysis of incompressible ocean waves.

\section{Analysis of Modes}\label{modeanalysis}

Let us define the following two dimensionless quantities:
\begin{equation}
\lambda = {\omega\over2\Omega}, \qquad
\delta = {gh_0\over4\Omega^2R^2}. \label{lambdadelta}
\end{equation}
Substituting equation (\ref{effeos}) in equations
(\ref{dyn1})--(\ref{dyn3}) and rewriting in terms of $\lambda$ and
$\delta$, we obtain the following set of equations for $\Sigma^{'}$,
$u$ and $v$:
\begin{eqnarray}
-i\lambda{\Sigma^{'}\over \Sigma_0} + {im\over2\Omega R} u
+{1\over2\Omega}{dv\over dy} &=& 0, \label{d1}\\ -i\lambda u +
2im\Omega RA\delta{\Sigma^{'}\over\Sigma_0} -{y\over R} v &=& 0,
\label{d2}\\ -i\lambda v + 2\Omega R^2A\delta {d\over
dy}\left({\Sigma^{'}\over\Sigma_0}\right) +{y\over R} u &=&
0. \label{d3}
\end{eqnarray}
Equations (\ref{d1}) and (\ref{d2}) can be solved for $\Sigma^{'}$
and $u$ in terms of $v$ and $dv/dy$:
\begin{eqnarray}
{\Sigma^{'}\over\Sigma_0} &=& {i\over(\lambda^2-m^2A\delta)}
\left[{m\over 2\Omega R}\,{y\over R}\,v - {\lambda\over2\Omega}\,{dv\over
dy}\right], \label{Sigmap}\\ u &=& {i\over(\lambda^2-m^2A\delta)}
\left[\lambda\,{y\over R}\,v - mA\delta R\,{dv\over dy}\right]. \label{u}
\end{eqnarray}
Substituting these in equation (\ref{d2}), we obtain the following
second-order differential equation for $v$:
\begin{equation}
{d^2v\over dy^2} + \left[\left({\lambda^2\over A\delta} -
{m\over\lambda} -m^2\right) -{1\over A\delta}
{y^2\over R^2}\right] {v\over R^2} = 0. \label{vdifeq}
\end{equation}
Equation (\ref{vdifeq}) may be directly compared to equation (8.2) in
\citet{LH68}, noting that our $m$ and $A\delta$ are called $s$ and
$1/\epsilon$ in that paper.  The term $-m^2$ in our equation is not
present in \citet{LH68}; however, this term is negligible whenever the
assumption of an equatorially concentrated mode (on which our analysis
is based) is valid.

Before proceeding to solve the mode equations, we note that we are not
interested in modes with $m=0$, since they will not produce any
observable oscillations.  Nor are we interested in modes with very
large values of $|m|$ or modes with many nodes in the $y$-direction,
since these will again have little detectable variations in the
observed flux.  In the following, we limit our discussion to the
lowest order eigenfunctions in the $y$-direction.  The generalization
to higher order modes is, of course, trivial.

Equation (\ref{vdifeq}) is of standard form and its solutions are
Hermite functions.  The lowest order eigenfunction is given by
\begin{equation}
v(y) = 2C \Omega R \delta^{1/2}\, \exp\left[-{y^2\over
2(A\delta)^{1/2}R^2}\right],
\label{v1}
\end{equation}
where $C$ is an arbitrary dimensionless constant that measures the
mode amplitude, and we have introduced the factor $2\Omega R
\delta^{1/2}$ to scale the velocity by the sound speed.  The function
(\ref{v1}) is a solution of equation (\ref{vdifeq}) provided the
eigenvalue $\lambda$ satisfies the following quadratic equation
\begin{equation}
\lambda^2 -m(A\delta)^{1/2}\lambda -(A\delta)^{1/2} = 0.
\label{disp1}
\end{equation}
This gives two roots, which correspond to the $g$-modes discussed in
\S~4.1.  Actually, there is a third root, $\lambda=-m(A\delta)^{1/2}$,
but it is spurious since it causes the denominators in equations
(\ref{Sigmap}) and (\ref{u}) to vanish.  This spurious root is
replaced by another root, which we discuss in \S~4.3.

In order for the solution (\ref{v1}) to be physically valid it must
decay rapidly with increasing $|y|$.  This means that, in taking the
square root $(A\delta)^{1/2}$, we must select the solution with a
positive real part.  Further, in deriving the equations, we assumed
that the mode is concentrated around the equator.  For this to be
true, we see from equation (\ref{v1}) that we require the modulus of
$(A\delta)^{1/2}$ to be $\ll1$.  The quantity $A\sim1$, so we require
$\delta^{1/2} \ll 1$.  For a typical neutron star ($M\sim1.4M_\odot$,
$R\sim10$ km), $g\sim 2\times10^{14} {\rm \,cm\,s^{-2}}$, $h_0 \sim
200$ cm and $\Omega/2\pi \sim 500$ Hz.  We then find that
$\delta^{1/2} \sim 10^{-1.5}$, which is indeed small.  The mode is
thus restricted to $|y|/R \lesssim 0.2$, which is sufficiently
concentrated around the equator for our approximations to be valid.

The next-order solution of equation (\ref{vdifeq}) is
\begin{equation}
v(y) = 2C \Omega R\delta^{1/2}\, {y\over
(A\delta)^{1/4}R}\,\exp\left[-{y^2\over 2(A\delta)^{1/2}R^2}\right].
\label{v2}
\end{equation}
We have included an extra factor of $(A\delta)^{-1/4}$ to allow for
the typical value of $|y|/R$, and so $C$ is again a measure of the
dimensionless mode amplitude.  This solution gives the following
condition on the eigenvalue $\lambda$:
\begin{equation}
{\lambda^3\over A\delta} -{m\over\lambda} -{3\over (A\delta)^{1/2}}-m^2 =
0.
\label{disp2}
\end{equation}

Longuet-Higgins (1968, see also Heyl 2004) has shown that the surface
modes of a rotating sphere separate into three kinds: $g$-modes, $r$-modes
and Kelvin modes.  We discuss each of these below, focusing on the
lowest order eigenfunction in each case.

\subsection{$g$-Modes}

The lowest order $g$-modes correspond to the solution (\ref{v1}), with
the dimensionless mode frequency $\lambda$ obtained by solving
equation (\ref{disp1}),
\begin{equation}
\lambda \approx \pm (A\delta)^{1/4},
\label{glambda}
\end{equation}
where we have made use of the fact that $\delta$ is small.  For a
given $m$ there are two modes, one of which moves towards the east
(the mode with the $+$ sign) and the other moves towards the west (the
mode with the $-$ sign).  Using the typical numerical values given
above and setting $A\sim1$, we find that the typical mode frequency is
$\sim 200$ Hz, which is fairly large.  The eigenfunction has the form
(we show only the leading terms)
\begin{eqnarray}
{\Sigma^{'}(y)\over\Sigma_0} &\approx& \pm {iC\over A^{1/2}}\,{y\over
(A\delta)^{1/4}R}\,\exp\left[-{y^2\over 2(A\delta)^{1/2}R^2}\right],
\\
\frac{u(y)}{2\Omega R \delta^{1/2}} &\approx& \pm iC \frac{y}{
(A\delta)^{1/4}R}\,\exp\left[-{y^2\over 2(A\delta)^{1/2}R^2}\right],
\\
\frac{v(y)}{2\Omega R \delta^{1/2}} &=& C \exp\left[-{y^2\over
2(A\delta)^{1/2}R^2}\right].
\end{eqnarray}

The $g$-modes have a lower growth rate than the other two modes
discussed below.  For this reason, and also because they have larger
frequencies, these modes are probably not of great interest for burst
oscillations.

\subsection{$r$-Modes}

The lowest order $r$-modes are given by the solution (\ref{v2}), with
the eigenvalue obtained by solving the cubic dispersion relation
(\ref{disp2}).  Two of the roots correspond to higher order $g$-modes
and we ignore them.  The third root gives the $r$-mode.  In the limit
$(A\delta)^{1/2} \ll 1$, this root is approximately equal to
\begin{equation}
\lambda \approx -{1\over3}m(A\delta)^{1/2}.
\label{Rlambda}
\end{equation}
For $m=1,$ 2, the typical mode frequencies are $\sim 10$, 20 Hz,
respectively, The negative sign on $\lambda$ indicates that the
$r$-mode always travels towards the west.  Both of these properties
are just what are required for explaining burst oscillations
\citep{H04}.

The eigenfunction of the $r$-mode is given by
\begin{eqnarray}
{\Sigma^{'}(y)\over\Sigma_0} &\approx& -{3iC\over8mA^{1/2}(A\delta)^{1/4}}\, \left[1+
{2y^2\over (A\delta)^{1/2}R^2}\right]\, \exp\left[-{y^2\over
2(A\delta)^{1/2}R^2}\right],
\nonumber \\ \\
\frac{u(y)}{2 \Omega R \delta^{1/2}} &\approx& \frac{9iC}{8m(A\delta)^{1/4}}  \,\left[1-{2y^2\over3(A\delta)^{1/2}R^2}\right]\, \exp\left[-{y^2\over2(A\delta)^{1/2}R^2}\right],
\\
\frac{v(y)}{2\Omega R \delta^{1/2}} &=& C {y\over (A\delta)^{1/4}R}\, \exp\left[-{y^2\over 2(A\delta)^{1/2}R^2}\right].
\end{eqnarray}

\subsection{Kelvin Modes}

The Kelvin mode has $v$ identically equal to zero, and so it cannot be
obtained by solving the differential equation (\ref{vdifeq}).
Instead, one has to go back to the primitive equations
(\ref{d1})--(\ref{d2}).  The eigenvalue of this mode is
\begin{equation}
\lambda = m(A\delta)^{1/2},
\label{klambda}
\end{equation}
and the two nonzero components of the eigenfunction are
\begin{eqnarray}
{\Sigma^{'}(y)\over\Sigma_0} &=& \frac{C}{A^{1/2}} \exp\left[-{y^2\over
2(A\delta)^{1/2}R^2}\right], \\
\frac{u(y)}{2\Omega R \delta^{1/2}} &=& C \exp\left[-{y^2\over
2(A\delta)^{1/2}R^2}\right].
\end{eqnarray}

The Kelvin mode always moves towards the east.  For a given $m$, it
has a frequency equal to three times that of the $r$-mode.  The motions
of the plasma are entirely in the east-west direction ($v=0$) whereas
they occur in both the east-west and north-south directions in the
$r$-mode.

\section{Growth Rate of Modes}\label{growthrate}

Equations (\ref{glambda}), (\ref{Rlambda}) and (\ref{klambda}) give
the frequencies of the lowest order $g$-, $r$- and Kelvin modes.  If A
is real, then all the frequencies are real and there is no mode
growth.  However, we showed in \S~3 that $A$ is in general complex.
Therefore, the modes will either grow or decay with time.

\subsection{Criterion for Mode Growth}

Let us focus on the $r$-mode and Kelvin mode for now.  Recall that
these modes have angular frequencies $\omega\sim10^2$ radians per
second.  In contrast, the cooling time of the accreted layer is of
order a second, so that the parameter $\Omega_c \sim 1
\,\mathrm{s}^{-1}$.  The parameter $\Omega_h$ is even smaller during
the decline of a burst.  We are thus permitted to treat
$\Omega_c/\omega$ and $\Omega_h/\omega$ in equation (\ref{A}) as small
quantities.  We can then Taylor expand (\ref{A}) to obtain
\begin{equation}
A^{1/2} \approx A_R^{1/2}
\left[1+i\left(\alpha{\Omega_h\over\omega}-\beta{\Omega_c\over\omega}\right)\right],
\label{rootA}
\end{equation}
where
\begin{eqnarray}
A_R^{1/2} &=& \left[{(2\gamma-1)\over2\gamma}\right]^{1/2}, \\
\alpha &=& \frac{(\gamma-1)}{2\gamma(2\gamma-1)}[\gamma+\eta+(\gamma-1)\nu], 
\label{alpha}\\
\beta &=& \frac{(\gamma-1)}{2\gamma(2\gamma-1)}[3\gamma-4-\xi-(\gamma-1)\zeta].
\label{beta}
\end{eqnarray}
As discussed earlier, we should pick the positive root for $A_R^{1/2}$
to ensure that the eigenfunction decays at large $y$.

Substituting (\ref{rootA}) in equation (\ref{Rlambda}), we find for the
frequency of the $r$-mode
\begin{equation}
\omega \approx -{2m\Omega\over3}(A_R\delta)^{1/2}
\left[1+i\left(\alpha{\Omega_h\over\omega}-\beta{\Omega_c\over\omega}\right)\right].
\end{equation}
Since the magnitude of the imaginary term on the right is very small,
we immediately obtain
\begin{equation}
\omega \approx -{2m\Omega\over3}(A_R\delta)^{1/2}
+i(\alpha\Omega_h-\beta\Omega_c).
\label{Rgrowth}
\end{equation}
In a similar manner, the frequency of the Kelvin mode is found by
substituting (\ref{rootA}) in (\ref{klambda}):
\begin{equation}
\omega \approx 2m\Omega(A_R\delta)^{1/2}
+i(\alpha\Omega_h-\beta\Omega_c).
\label{kgrowth}
\end{equation}
In the case of the $g$-mode, (\ref{glambda}) shows that $\lambda \propto
(A\delta)^{1/4}$ rather than $(A\delta)^{1/2}$.  Consequently, the
imaginary part of $\omega$  is reduced by a factor of 2:
\begin{equation}
\omega \approx \pm 2\Omega(A_R\delta)^{1/4}
+{i\over2}(\alpha\Omega_h-\beta\Omega_c).
\label{ggrowth}
\end{equation}

A mode is unstable if its $\omega$ has a positive imaginary part.
Thus, all three modes give the same condition for instability, viz.,
\begin{equation}
\alpha\Omega_h - \beta\Omega_c > 0,
\label{condition}
\end{equation}
with $\alpha$ and $\beta$ defined in equations (\ref{alpha}) and
(\ref{beta}).  Before applying this result to burst oscillations, we
first discuss the physics of the instability.

\subsection{Physics of the Instability}

\subsubsection{Local Thermal Stability}

We begin by deriving the standard criterion for thermal stability of a
burning layer.  We consider linear perturbations of the entropy
equation (\ref{entropy}), but we assume that there are no nonradial
perturbations, i.e., $\Sigma^{'} = 0$.  From equations (\ref{Teq}),
(\ref{entropypert}), (\ref{heatpert}), and (\ref{coolpert}), setting
$\Sigma^{'}=0$, we find
\begin{equation}
\frac{\gamma}{(\gamma-1)} \frac{\partial \ln T^{'}}{\partial t} = 
(\nu-\eta) \Omega_h - (4+\xi-\zeta)  \Omega_c.
\end{equation}
Clearly, the accreted layer is thermally unstable when the 
right side of the equation is positive.  Thus, the criterion for
thermal instability is
\begin{equation}
{\Omega_h\over\Omega_c} >  \frac{4 + \xi - \zeta}{\nu -\eta}.
\label{thermalstabeqn1}
\end{equation}

Consider the accreted layer on a neutron star prior to the onset of a
type I X-ray burst.  The layer undergoes steady thermonuclear burning at
some rate.  Steady state requires in general $\Sigma_0 \epsilon_0 =
F_0 - F_b$.  However, the flux $F_b$ entering the base of the accreted
layer is typically much less than the flux $F_0$ escaping from the
surface \citep[e.g.,][]{B00,B04}, so $F_0 \gg F_b$.  Therefore, we
expect $\Sigma_0\epsilon_0 \approx F_0$, which is equivalent to the
condition $\Omega_h \approx \Omega_c$.  The criterion for thermal
instability then becomes
\begin{equation}
\nu -\eta  >  4 + \xi - \zeta,
\label{thermalstabeqn2}
\end{equation}
which is equivalent to equation (23) of \citet{B98}.  

The thermal instability is driven by variations in $T$.  Consider a
positive temperature perturbation at constant pressure (the pressure
is constant because we have assumed hydrostatic equilibrium and a
constant $\Sigma$).  This leads to a corresponding decrease in $\rho$
according to equation (\ref{Teq}).  If the marginal increase in the
nuclear energy generation rate $\epsilon^{'} \propto (\nu-\eta)$
exceeds the marginal increase in the effective cooling rate of the
layer $F^{'}/\Sigma \propto (4 + \xi - \zeta)$, the temperature of the
accreted layer increases further and a thermonuclear runaway ensues.
This explains the form of the two factors on either side of equation
(\ref{thermalstabeqn2}) as well as the direction of the inequality.

The thermal instability is paramount for understanding the triggering
of type I X-ray bursts, and there is considerable literature on this subject
\citep{SH65,HvH75,FHM81,P83,FL87,B98,NH03,CN06a,CN06b,
FGWD06,FTGWC07}.  However, once a burst has been
initiated, the temperature of the burning matter increases dramatically
until the burning layer finds a new quasi-steady-state in which the
temperature is typically $\sim10^9$ K.  In this new state, the indices
$\nu$, $\eta$, $\xi$ and $\zeta$ have different values.  In
particular, the temperature index $\nu$ of the nuclear energy
generation rate $\epsilon$, which is large at the cooler temperatures
characteristic of burst ignition, now becomes much more modest.  As a
result, the condition (\ref{thermalstabeqn2}) is no longer satisfied,
and the layer is thermally stable.  Thus, as far as pure thermal
perturbations are concerned, the burning layer during a burst is
stable and the observed oscillations in the burst tail cannot be
attributed to thermal instability.

\subsubsection{Nonradial Mode Stability}

The instability that we have analyzed in this paper is associated with
nonradial surface modes, and it depends crucially on the fact that
$\Sigma$ is perturbed.  The physics is similar to that which causes
nonradial stellar pulsations in Cepheids and other variable stars
\citep{C80}.  The one big difference is that the primary driving
mechanism here is nuclear reactions (the so-called $\epsilon$
mechanism) rather than opacity (the $\kappa$ mechanism).  \citet{MT87} 
and \citet{SL96} investigated the driving of surface modes via the 
$\epsilon$ mechanism in thermally unstable stellar envelopes.  In contrast, 
we have investigated the driving of surface modes in thermally 
stable envelopes.  As \citet{PB04}
have emphasized, the stability criterion for nonradial
perturbations is distinct from the criterion for thermal stability.  

In stellar pulsation theory, the following simple criterion is used to
decide whether a particular layer of matter is a driving or damping
region for pulsations \citep{C80}: A region that gains heat when
compressed is a driving region, whereas a region that loses heat when
compressed is a damping region.  In the nonradial surface modes of
our problem, compression is induced by variations in $\Sigma$.  Since
we are considering a simple one-zone vertical structure for the 
layer, we merely have to determine whether the specific entropy of the
plasma increases or decreases when $\Sigma$ increases.  If the entropy
increases with increasing $\Sigma$ it means that the plasma gains heat
when it is compressed and the mode is unstable; if the entropy
decreases, the mode is stable.

In equilibrium, an unperturbed column satisfies $\Sigma_0\epsilon_0 =
(F_0-F_b)$.  Let us imagine compressing the column laterally so that
the surface density increases to $\Sigma_0+\Sigma^{'}$.  Let the
compressed column relax to a new thermal equilibrium state.  Assuming
$F_b$ is unchanged, we require $(\Sigma\epsilon)^{'} = F'$.
Substituting for these quantities from equations (\ref{heatpert}) and
(\ref{coolpert}), we can solve for $\rho'/\rho_0$ in terms of
$\Sigma^{'}/\Sigma_0$:
\begin{equation}
{\rho'\over\rho_0} = {-(\nu+1)\Omega_h + (3-\zeta)\Omega_c \over
(\eta-\nu)\Omega_h + (4+\xi-\zeta)\Omega_c}\,{\Sigma^{'}\over\Sigma_0}.
\end{equation}
From equation (\ref{Teq}), we can also obtain an expression for
$T'/T_0$ in terms of $\Sigma^{'}/\Sigma_0$:
\begin{equation}
{T'\over T_0} = {(\eta+1)\Omega_h + (\xi+1)\Omega_c \over
(\eta-\nu)\Omega_h + (4+\xi-\zeta)\Omega_c}\,{\Sigma^{'}\over\Sigma_0}.
\end{equation}
We may now use equation (\ref{entropy2}) to estimate the perturbation
in the specific entropy of the plasma:
\begin{equation}
{s'\over c_V} = {[\gamma+\eta+(\gamma-1)\nu]\Omega_h - 
[3\gamma-4-\xi-(\gamma-1)\zeta]\Omega_c \over
(\eta-\nu)\Omega_h + (4+\xi-\zeta)\Omega_c}\,{\Sigma^{'}\over\Sigma_0}.
\label{criterion}
\end{equation}

By the simple criterion mentioned earlier, the layer of plasma under
consideration will be a driving layer for pulsations if $s'$ increases
with increasing $\Sigma^{'}$, i.e., if the coefficient on the right
side of equation (\ref{criterion}) is positive.  During the burst, the
layer is thermally stable, and so by equation (\ref{thermalstabeqn1})
the denominator of this coefficient is positive. Thus, for nonradial
instability, we require the numerator also to be positive, i.e.,
\begin{equation}
[\gamma+\eta+(\gamma-1)\nu]\Omega_h -
[3\gamma-4-\xi-(\gamma-1)\zeta]\Omega_c > 0,
\label{condition2}
\end{equation}
which is identical to equation (\ref{condition}).

The analysis presented here assumes that the column comes into thermal
equilibrium after it is compressed.  This is not a good approximation
in the case of a mode, especially if the mode frequency is much larger
than the heating/cooling rate, as in our problem.  Nevertheless, it is
easy to see that the same criterion for instability must hold.
Consider a column of matter that undergoes a positive $\Sigma$
perturbation as a result of lateral compression induced by an
oscillating mode.  Since the compressed column is not in thermal
equilibrium, its specific entropy will begin to increase.  Of course,
the entropy will not have time to reach its equilibrium value since
the mode varies rapidly.  Nevertheless, by the time the compression
begins to reduce, the entropy of the plasma will be a little larger than
before the column was compressed.  The increased entropy will be
reflected in an increased effective pressure $P$, and so the expansion
will be a little more energetic compared to the initial compression.
As a result, the amplitude of the next half-cycle of the wave will be
a little larger.  The process is obviously self-sustaining and the
wave amplitude will grow exponentially with time.

\subsection{Application to Type I X-ray Bursts}\label{burstapp}

Surface modes during a type I X-ray burst will be unstable whenever
the condition (\ref{condition}) or (\ref{condition2}) is satisfied.
At the high temperatures that are present during a burst, the opacity
scales according to the approximate formulae given in \citet{Pre83},
which give $-0.1 \lesssim \xi \lesssim 0$, $-0.5 \lesssim \zeta
\lesssim 0$.  We assume $\xi=0$, $\zeta=-0.25$.  During the decline
phase of a burst, the escaping flux is obviously larger than the rate
at which heat is added via nuclear reactions, i.e., $\Omega_c >
\Omega_h$.  The ratio of the two rates, however, depends on details
which are beyond the scope of this work.

When gas pressure dominates, we have $\gamma=5/3$, and the condition
for nonradial instability becomes
\begin{equation}
\frac{\Omega_h}{\Omega_c} > \frac{7}{10 + 6 \eta + 4\nu}.
\label{criterion2}
\end{equation}
On the other hand, when radiation pressure dominates, $\gamma=4/3$,
and the condition is
\begin{equation}
\frac{\Omega_h}{\Omega_c} > \frac{1}{16 + 12 \eta + 4\nu}.
\label{criterion3}
\end{equation}
Whether or not these conditions are satisfied depends on the kind of
burning that takes place during the burst.

\subsubsection{Helium Burning}

The nuclear flow during a helium-rich burst includes few weak
interactions, and thus nuclear burning is rapid and occurs over a
relatively short timescale.  Consequently, $\Omega_h$ is likely to be
a considerable fraction of $\Omega_c$ during the burst.  Furthermore,
if heating during the burst is primarily through helium-burning
reactions, we expect $\epsilon$ to depend on both density and
temperature.  Depending on whether the triple-$\alpha$ reaction or
some other (two-particle) $\alpha$ chain reaction is important,
$\eta=2$ or 1.  As for $\nu$, we expect it to be of order a few.

The temperatures achieved during a burst can often exceed $10^{9}$ K,
in which case radiation pressure becomes important
\citep{J77,Wetal04}.  This is especially true of really powerful,
e.g., photospheric radius expansion (PRE), bursts.  If radiation
pressure dominates, the instability criterion is given by
(\ref{criterion3}), which is easily satisfied even if nuclear burning
supplies only a few percent of the cooling flux.  Thus, instability is
very likely in such bursts.  Observationally, a good number of bursts
with oscillations are indeed found to be PRE bursts
\citep{MFMB00,Getal07}.  PRE bursts are typically helium-rich, and thus we
conclude that helium burning is conducive to burst oscillations.

If gas pressure dominates, then the instability criterion is given by
(\ref{criterion2}), which for the typical values of $\eta$ and $\nu$
given above, will be satisfied so long as $\Omega_h$ is greater than
or of order $\Omega_c/4$.  The exact amount of nuclear burning during
the decay phase of a helium burst is unclear.  
This issue is complicated by the fact that, although observations indicate 
that short, helium-dominated bursts occur at high accretion rates, all 
current detailed numerical burst models predict that such bursts occur 
only at low accretion rates.  However, \citet{Wetal04} show that 
nuclear burning is substantial for the majority of the duration of 
these bursts.  Thus, it is probably not unreasonable to expect
helium-rich bursts to have oscillations even when they are gas-pressure
dominated.

Helium burning is typically rapid, and so bursts in which
helium burning dominates are relatively short, no more than several
seconds.  Hence we suggest that short bright bursts are the best
places to look for burst oscillations.

Given the short duration of these bursts, one issue is whether there
will be enough time for the mode to grow to a nonlinear amplitude.
The growth rate is given by equation (\ref{ggrowth}) and is generally
a fraction of the heating/cooling rate (since $\alpha$, $\beta$ are
typically less than unity).  The burst duration, on the other hand, is
a factor of several longer than the heating/cooling time scale.  Thus
the growth time is expected to be comparable to the burst duration.
It is then a delicate question whether or not the oscillations will
have enough time to grow to a large amplitude before the burst dies
away.  This can be decided only with detailed calculations.

\subsubsection{Hydrogen Burning}

In hydrogen-rich bursts, much of the heating is supplied by hydrogen
burning via the $rp$-process of \citet{WW81}.  The nuclear energy
generation rate is limited by a long series of slow $\beta$-decays
which act as bottlenecks in the nuclear flow.  Because of this, the
reaction rate per unit mass is less sensitive to density and
temperature, i.e., $\eta$ and $\nu$ are closer to zero, compared to
the case of helium burning.  This means that $\Omega_h/\Omega_c$ has
to be larger than in the case of helium-burning bursts before the
instability will be triggered.  Furthermore, because of the slow 
$rp$-process, these bursts are not very bright at the peak and are unlikely
to become very radiation dominated.  As a result, they need to satisfy
the more stringent criterion (\ref{criterion2}) for instability.  For
all these reasons, we do not expect these bursts to exhibit
oscillations.

Hydrogen-rich bursts usually last substantially longer than
helium-burning bursts because of the slow $rp$-process.
Observationally, we thus suggest that long-duration bursts will in
general show fewer episodes of oscillations.

\subsubsection{Oscillations During Burst Rise}

As explained in \S1, oscillations during the burst rise are
interpreted in terms of a non-axisymmetric burning spot associated
with the initial burst trigger.  However, unstable nonradial modes of
the kind discussed in this paper should also be considered.  During
the rise, clearly $\Omega_h > \Omega_c$, and so mode instability is
quite likely.  Indeed, \citet{MCGP02} found that bursts from which 
oscillations are detected during the rising phase show the largest 
frequency drifts, and they concluded that the mechanism that generates 
oscillations in the burst tail begins with the initial burst 
trigger.  Perhaps one could interpret the propagating spot model
of \citet{SLU02} as a nonlinear traveling mode.  On the other hand,
the burst rise is rather rapid, and it is possible that there is not
enough time for a mode to grow to saturation amplitude.

\subsubsection{Amplitude of the Oscillations}

An unstable mode is likely to saturate at a dimensionless amplitude $C
\lesssim 1$.  Thus we expect $\Sigma^{'}/\Sigma_0 \lesssim 1$ and, by
equation (\ref{coolpert}), the amplitude of variation of the local
escaping flux $F^{'}$ to also be of this order.  In the case of
$\beta$ Cephei stars, temperature fluctuations in nonradial
pulsational modes are typically about $\pm10\%$ (Berdyugina et
al. 2003), which correspond to flux variations of $\pm50\%$.  Let us
assume a similar amplitude for $F^{'}$ in our problem.  Since the
nonradial modes we have considered occupy only about $20\%$ of the
surface of the star (see the discussion in \S \ref{modeanalysis}), the
observed flux variations in burst oscillations are expected to be
$\sim 10\%$.  This estimate is in accord with the observed amplitudes,
which are $\lesssim 15\%$ \citep{MOC02,SB06}.

\subsubsection{Radial Structure of the Mode}

The vertical thickness of the burning layer varies as in equation
(\ref{h}).  Thus $h'/h_0$ is of order $\Sigma'/\Sigma_0$, which itself
is roughly the transverse displacement divided by the the wavelength
of the mode.  The latter quantity is of order $h_0/R$, so we expect
the ratio of the radial displacement to the transverse displacement to
be of order $h_0/R \sim 10^{-4}$.  This value is consistent with the
equivalent ratio $\xi_r/\xi_{\perp}$ of \citet{PB04}.  Figure 4 of
\citet{PB04} illustrates that $\xi_{\perp}$ is roughly constant within
the burning layer and drops rapidly below the burning layer.  In our
simple model, $\xi_{\perp}$ is in effect assumed to be constant
throughout the burning layer and to go to zero immediately below the
layer.  Our approximation is thus reasonably consistent with the more
detailed model of \citet{PB04}.

\subsubsection{Frequency Drift}

A general feature of burst oscillations is that the frequency drifts
upward during the course of a burst, typically by about 1 to 4 Hz
(e.g., Muno et al. 2003; Piro \& Bildsten 2005b).  Heyl (2004) argued
that this is the result of the mode frequency in the rotating frame
(the real part of $\omega$) decreasing during the course of the burst
as the plasma cools.  For a backward moving mode such as the $r$-mode,
there would be a corresponding increase in the observed oscillation
frequency.

For an $m=1$ $r$-mode, the typical value of 
$\mathrm{Re}\, (\omega/2\pi) \sim 10\,
(T/10^9\,{\rm K})^{1/2}$ Hz (Piro \& Bildsten 2005b).  As a burst decays
from its peak, the temperature is expected to vary by about a factor
of 2, from say $10^9$ K to $5\times10^8$ K, and the corresponding
change in $\mathrm{Re}\, (\omega/2\pi)$ is expected to be about 3~Hz.  
This estimate is
comparable to the upper end of the range of observed frequency drifts,
but is perhaps a little too large to be consistent with the smaller
drifts $\sim1$~Hz seen in some systems.  To solve this problem (though
we are not convinced that there is indeed a serious problem), Heyl
(2004) suggested that the relevant modes may be some kind of
photospheric disturbances, while Piro \& Bildsten (2005b) proposed a
transition from a surface wave (of the sort described in the present
paper) to a crustal interface mode.

\section{Discussion}\label{discussion}

As mentioned in \S \ref{introduction}, oscillations during the decay
phase of type I X-ray bursts are detected predominantly when the
inferred accretion rate is high.  In \S \ref{burstapp} we deduced that
oscillations should occur only in short, helium-dominated bursts.
Observations indicate that the burst decay timescale decreases with
increasing $\dot{M}$ \citep{vPPL88,Cetal03,Getal07}.  Therefore, we
suggest that oscillations are detected predominantly at high accretion
rates primarily because the burst durations are lower at these rates.
To our knowledge, all bursts from neutron stars that are not
accretion-powered pulsars and that exhibit oscillations in their tails
have short durations.

We stress that our work may not apply to accretion-powered millisecond
pulsars from which burst oscillations have been detected.  There are
several differences between bursts from pulsars and those from
non-pulsars \citep[see, e.g.,][]{PB05b,WS06}.  In particular,
oscillations have been detected only in long bursts from the two
pulsars that exhibit burst oscillations, XTE J1814--338
\citep{SMSintZ03} and SAX J1808.4--3658 \citep{intZetal01}.  Thus, we
suspect that a different mechanism generates burst oscillations in
accretion-powered millisecond pulsars \citep[see, e.g.,][]{LKR07}.

There are several issues that we have not addressed.

First, we have not attempted to calculate the crucial
heating-to-cooling ratio, $\Omega_h/\Omega_c$, which figures
prominently in our instability criterion (\ref{condition}),
(\ref{condition2}).  This can be done only with detailed multi-zone
time-dependent burst simulations \citep[e.g.,][]{Wetal04,FGWD06}.

Second, we have only carried out a linear stability analysis, and 
we had to use a plausibility argument to estimate the mode amplitude at
saturation and the amplitude of the observed flux variations.  To
obtain a more accurate estimate, one needs to carry out full
nonlinear calculations, perhaps along the lines of \citet{SLU02}.

Third, it appears from equations (\ref{Rgrowth}) and (\ref{kgrowth})
that the $r$-mode and Kelvin mode have identical growth rates and
should therefore both be observed.  Also, the linear growth rate does
not depend on the azimuthal mode number $m$.  However, the observed
oscillations seem to consist predominantly of the $r$-mode
\citep{H04} with $m=1$.  Differences between the modes might emerge in
the nonlinear regime, driven perhaps by the fact that the frequency of
the $r$-mode is a third that of the Kelvin mode and the spatial scale
of the $m=1$ mode is larger than that of higher $m$ modes, but it is
mere speculation at this point.

Finally, oscillations have been detected in the decay phase of one
superburst.  We suspect that a mechanism similar to the one we have
described for helium-burning type I X-ray bursts generates these
oscillations as well.  However, unlike normal helium-triggered bursts,
the physics of the ignition and subsequent nuclear burning of
superbursts are not well understood \citep[see][for some recent work
on this topic]{WBB06}, so we cannot make a definitive statement
regarding superburst oscillations.

\acknowledgments

We would like to thank Tony Piro for his critical reading of an
earlier draft of this manuscript, Lars Bildsten for helpful
discussions, and the referee for thoughtful suggestions that helped us 
improve the quality of this work.  This work was supported by NASA grant 
NNG04GL38G.  R. C. was supported in part by the National Science 
Foundation under Grant No. PHY99-07949.

\bibliographystyle{apj}

\begin{thebibliography}
\expandafter\ifx\csname natexlab\endcsname\relax\def\natexlab#1{#1}\fi

\bibitem[{{Babushkina} {et~al.}(1975){Babushkina}, {Bratolyubova-Tsulukidze},
  {Kudryavtsev}, {Melioranskiy}, {Savenko}, \& {Yushkov}}]{Betal75}
{Babushkina}, O.~P., {Bratolyubova-Tsulukidze}, L.~S., {Kudryavtsev}, M.~I.,
  {Melioranskiy}, A.~S., {Savenko}, I.~A., \& {Yushkov}, B.~Y. 1975, Soviet
  Astronomy Letters, 1, 32

\bibitem[{{Belian} {et~al.}(1976){Belian}, {Conner}, \& {Evans}}]{BCE76}
{Belian}, R.~D., {Conner}, J.~P., \& {Evans}, W.~D. 1976, \apjl, 206, L135

\bibitem[Berdyugina et al.(2003)]{Berdyugina}
Berdyugina, S. V., Telting, J. H., Korhonen, H., \& Schrijvers, C.
2003, A\&A, 406, 281

\bibitem[{{Bhattacharyya}(2007)}]{B07} {Bhattacharyya}, S.\ 2007, 
\mnras, 377, 198

\bibitem[{{Bhattacharyya} {et~al.}(2006){Bhattacharyya}, {Strohmayer},
  {Markwardt}, \& {Swank}}]{BSMS06}
{Bhattacharyya}, S., {Strohmayer}, T.~E., {Markwardt}, C.~B., \& {Swank}, J.~H.
  2006, \apjl, 639, L31

\bibitem[{{Bildsten}(1995)}]{B95}
{Bildsten}, L. 1995, \apj, 438, 852

\bibitem[{{Bildsten}(1998)}]{B98}
{Bildsten}, L. 1998, in The Many Faces of Neutron Stars, ed. R. Buccheri, J.
  van Paradijs, \& M. A. Alpar (NATO ASI Ser. C, 515; Dordrecht: Kluwer), 419

\bibitem[Bildsten et al.(1996)]{BUC96} Bildsten, L., 
Ushomirsky, G., \& Cutler, C.\ 1996, \apj, 460, 827 

\bibitem[{{Brown}(2000)}]{B00}
{Brown}, E.~F. 2000, \apj, 531, 988

\bibitem[{{Brown}(2004)}]{B04}
---. 2004, \apjl, 614, L57

\bibitem[Cooper \& Narayan(2006a)]{CN06a} Cooper, R.~L. \& 
Narayan, R. 2006a, \apjl, 648, L123

\bibitem[Cooper \& Narayan(2006b)]{CN06b} ---. 2006b, \apj, 652, 584 

\bibitem[{{Cooper} \& {Narayan}(2007)}]{CN07} ---. 2007, \apj, 657, L29

\bibitem[{{Cornelisse} {et~al.}(2003){Cornelisse}, {in't Zand}, {Verbunt},
  {Kuulkers}, {Heise}, {den Hartog}, {Cocchi}, {Natalucci}, {Bazzano}, \&
  {Ubertini}}]{Cetal03}
{Cornelisse}, R., {in't Zand}, J.~J.~M., {Verbunt}, F., {Kuulkers}, E.,
  {Heise}, J., {den Hartog}, P.~R., {Cocchi}, M., {Natalucci}, L., {Bazzano},
  A., \& {Ubertini}, P. 2003, \aap, 405, 1033

\bibitem[{{Cox}(1980){Cox}}]{C80}
Cox, J. P. 1980, Theory of Stellar Pulsation, Princeton Univ. Press

\bibitem[{{Cumming}(2004)}]{C04}
{Cumming}, A. 2004, Nuclear Physics B Proceedings Supplements, 132, 435

\bibitem[{{Cumming}(2005)}]{C05b} ---. 2005, \apj, 630, 441

\bibitem[Fisker et al.(2006)]{FGWD06} Fisker, J.~L., 
G{\"o}rres, J., Wiescher, M., \& Davids, B. 2006, \apj, 650, 332

\bibitem[Fisker et al.(2007)]{FTGWC07} Fisker, J.~L., 
Tan, W., G{\"o}rres, J., Wiescher, M., \& Cooper, R.~L.\ 2007, \apj, 
submitted (astro-ph/0702412) 

\bibitem[{{Franco}(2001)}]{F01}
{Franco}, L.~M. 2001, \apj, 554, 340

\bibitem[{{Fryxell} \& {Woosley}(1982)}]{FW82}
{Fryxell}, B.~A. \& {Woosley}, S.~E. 1982, \apj, 261, 332

\bibitem[{{Fujimoto} {et~al.}(1981){Fujimoto}, {Hanawa}, \& {Miyaji}}]{FHM81}
{Fujimoto}, M.~Y., {Hanawa}, T., \& {Miyaji}, S. 1981, \apj, 247, 267

\bibitem[Fushiki \& Lamb(1987)]{FL87} Fushiki, I. \& Lamb, 
D.~Q.\ 1987, \apjl, 323, L55

\bibitem[{{Galloway} {et~al.}(2007){Galloway}, {Muno}, {Hartman}, {Savov},
  {Psaltis}, \& {Chakrabarty}}]{Getal07}
{Galloway}, D.~K., {Muno}, M.~P., {Hartman}, J.~M., {Savov}, P., {Psaltis}, D.,
  \& {Chakrabarty}, D. 2007, ApJS, submitted (astro-ph/0608259)

\bibitem[{{Giles} {et~al.}(2002){Giles}, {Hill}, {Strohmayer}, \&
  {Cummings}}]{GHSC02}
{Giles}, A.~B., {Hill}, K.~M., {Strohmayer}, T.~E., \& {Cummings}, N. 2002,
  \apj, 568, 279

\bibitem[{{Grindlay} {et~al.}(1976){Grindlay}, {Gursky}, {Schnopper},
  {Parsignault}, {Heise}, {Brinkman}, \& {Schrijver}}]{Getal76}
{Grindlay}, J., {Gursky}, H., {Schnopper}, H., {Parsignault}, D.~R., {Heise},
  J., {Brinkman}, A.~C., \& {Schrijver}, J. 1976, \apjl, 205, L127

\bibitem[{{Grindlay} \& {Heise}(1975)}]{GH75}
{Grindlay}, J. \& {Heise}, J. 1975, \iaucirc, 2879

\bibitem[Hansen \& van Horn(1975)]{HvH75} Hansen, C.~J. \& 
van Horn, H.~M. 1975, \apj, 195, 735 

\bibitem[{{Heyl}(2004)}]{H04}
{Heyl}, J.~S. 2004, \apj, 600, 939

\bibitem[{{Heyl}(2005)}]{H05}
---. 2005, \mnras, 361, 504

\bibitem[{{Houghton}(2002)}]{H02}
{Houghton}, J.~T. 2002, {The Physics of Atmospheres} (3rd ed.; Cambridge:
  Cambridge Univ. Press)

\bibitem[{{in't Zand} {et~al.}(2001){in't Zand}, {Cornelisse}, {Kuulkers},
  {Heise}, {Kuiper}, {Bazzano}, {Cocchi}, {Muller}, {Natalucci}, {Smith}, \&
  {Ubertini}}]{intZetal01}
{in't Zand}, J.~J.~M., {Cornelisse}, R., {Kuulkers}, E., {Heise}, J., {Kuiper},
  L., {Bazzano}, A., {Cocchi}, M., {Muller}, J.~M., {Natalucci}, L., {Smith},
  M.~J.~S., \& {Ubertini}, P. 2001, \aap, 372, 916

\bibitem[{{Joss}(1977)}]{J77}
{Joss}, P.~C. 1977, \nat, 270, 310

\bibitem[{{Joss}(1978)}]{J78}
---. 1978, \apjl, 225, L123

\bibitem[{{Kaaret} {et~al.}(2007){Kaaret}, {Prieskorn}, {in 't Zand}, {Brandt},
  {Lund}, {Mereghetti}, {Gotz}, {Kuulkers}, \& {Tomsick}}]{Ketal07}
{Kaaret}, P., {Prieskorn}, Z., {in 't Zand}, J.~J.~M., {Brandt}, S., {Lund},
  N., {Mereghetti}, S., {Gotz}, D., {Kuulkers}, E., \& {Tomsick}, J.~A. 2007,
  \apjl, 657, L97

\bibitem[{{Lamb} \& {Lamb}(1977)}]{LL77}
{Lamb}, D.~Q. \& {Lamb}, F.~K. 1977, in Eight Texas Symposium on Relativistic
  Astrophysics, ed. M.~D. {Papagiannis} (New York: New York Academy of
  Sciences), 261

\bibitem[{{Lamb} \& {Lamb}(1978)}]{LL78}
{Lamb}, D.~Q. \& {Lamb}, F.~K. 1978, \apj, 220, 291

\bibitem[{{Lee}(2004)}]{L04}
{Lee}, U. 2004, \apj, 600, 914

\bibitem[{{Lee} \& {Strohmayer}(2005)}]{LS05}
{Lee}, U. \& {Strohmayer}, T.~E. 2005, \mnras, 361, 659

\bibitem[{{Longuet-Higgins}(1968)}]{LH68}
{Longuet-Higgins}, M.~S. 1968, Philos. Trans. R. Soc. London A, 262, 511

\bibitem[{{Lovelace} {et~al.}(2007){Lovelace}, {Kulkarni}, \&
  {Romanova}}]{LKR07}
{Lovelace}, R.~V.~E., {Kulkarni}, A.~K., \& {Romanova}, M.~M. 2007, \apj, 656,
  393

\bibitem[{{Maraschi} \& {Cavaliere}(1977)}]{MC77}
{Maraschi}, L. \& {Cavaliere}, A. 1977, Highlights of Astronomy, 4, 127

\bibitem[{{McDermott} \& {Taam}(1987)}]{MT87}
{McDermott}, P.~N. \& {Taam}, R.~E. 1987, \apj, 318, 278

\bibitem[{{Muno} {et~al.}(2002){Muno}, {Chakrabarty}, {Galloway}, \&
  {Psaltis}}]{MCGP02}
{Muno}, M.~P., {Chakrabarty}, D., {Galloway}, D.~K., \& {Psaltis}, D. 2002,
  \apj, 580, 1048

\bibitem[{{Muno} {et~al.}(2000){Muno}, {Fox}, {Morgan}, \& {Bildsten}}]{MFMB00}
{Muno}, M.~P., {Fox}, D.~W., {Morgan}, E.~H., \& {Bildsten}, L. 2000, \apj,
  542, 1016

\bibitem[{{Muno} {et~al.}(2004){Muno}, {Galloway}, \& {Chakrabarty}}]{MGC04}
{Muno}, M.~P., {Galloway}, D.~K., \& {Chakrabarty}, D. 2004, \apj, 608, 930

\bibitem[{{Muno}, {\"Ozel}, \&
{Chakrabarty}(2002)}]{MOC02} {Muno}, M.~P., {\"Ozel}, F., \& {Chakrabarty},
D. 2002, \apj, 581, 550

\bibitem[Narayan \& Heyl(2003)]{NH03} Narayan, R. \& Heyl, 
J.~S.\ 2003, \apj, 599, 419

\bibitem[Paczy{\'n}ski(1983a)]{P83} Paczy{\'n}ski, B.\ 1983a, \apj, 
264, 282 

\bibitem[{{Paczy{\'n}ski}(1983b)}]{Pre83} 
---. 1983b, \apj, 267, 315

\bibitem[{{Piro} \& {Bildsten}(2004)}]{PB04}
{Piro}, A.~L. \& {Bildsten}, L. 2004, \apj, 603, 252

\bibitem[{{Piro} \& {Bildsten}(2005a)}]{PB05a}
---. 2005a, \apj, 619, 1054 

\bibitem[{{Piro} \& {Bildsten}(2005b)}]{PB05b}
---. 2005b, \apj, 629, 438

\bibitem[{{Piro} \& {Bildsten}(2006)}]{PB06}
---. 2006, \apj, 638, 968

\bibitem[{{Schwarzschild} \& {H{\"a}rm}(1965)}]{SH65} 
Schwarzschild, M. \& H{\"a}rm, R. 1965, \apj, 142, 855 

\bibitem[{{Shara}(1982)}]{S82}
{Shara}, M.~M. 1982, \apj, 261, 649

\bibitem[{{Spitkovsky} {et~al.}(2002){Spitkovsky}, {Levin}, \&
  {Ushomirsky}}]{SLU02}
{Spitkovsky}, A., {Levin}, Y., \& {Ushomirsky}, G. 2002, \apj, 566, 1018

\bibitem[{{Strohmayer} \& {Bildsten}(2006)}]{SB06}
{Strohmayer}, T. \& {Bildsten}, L. 2006, in Compact Stellar X-Ray Sources, ed.
  W.~H.~G. {Lewin and M. van der Klis (Cambridge: Cambridge Univ. Press)}, 113

\bibitem[{{Strohmayer} \& {Lee}(1996)}]{SL96} {Strohmayer}, T.~E., 
\& {Lee}, U.\ 1996, \apj, 467, 773

\bibitem[{{Strohmayer} \& {Markwardt}(1999)}]{SM99}
{Strohmayer}, T.~E. \& {Markwardt}, C.~B. 1999, \apjl, 516, L81

\bibitem[{{Strohmayer} \& {Markwardt}(2002)}]{SM02}
---. 2002, \apj, 577, 337

\bibitem[{{Strohmayer} {et~al.}(2003){Strohmayer}, {Markwardt}, {Swank}, \&
  {in't Zand}}]{SMSintZ03}
{Strohmayer}, T.~E., {Markwardt}, C.~B., {Swank}, J.~H., \& {in't Zand}, J.
  2003, \apjl, 596, L67

\bibitem[{{Strohmayer} {et~al.}(1997){Strohmayer}, {Zhang}, \& {Swank}}]{SZS97}
{Strohmayer}, T.~E., {Zhang}, W., \& {Swank}, J.~H. 1997, \apjl, 487, L77

\bibitem[{{Strohmayer} {et~al.}(1996){Strohmayer}, {Zhang}, {Swank}, {Smale},
  {Titarchuk}, {Day}, \& {Lee}}]{Setal96}
{Strohmayer}, T.~E., {Zhang}, W., {Swank}, J.~H., {Smale}, A., {Titarchuk}, L.,
  {Day}, C., \& {Lee}, U. 1996, \apjl, 469, L9

\bibitem[{{Strohmayer} {et~al.}(1998){Strohmayer}, {Zhang}, {Swank}, {White},
  \& {Lapidus}}]{SZSWL98}
{Strohmayer}, T.~E., {Zhang}, W., {Swank}, J.~H., {White}, N.~E., \& {Lapidus},
  I. 1998, \apjl, 498, L135

\bibitem[{{van Paradijs} {et~al.}(1988){van Paradijs}, {Penninx}, \&
  {Lewin}}]{vPPL88}
{van Paradijs}, J., {Penninx}, W., \& {Lewin}, W.~H.~G. 1988, \mnras, 233, 437

\bibitem[{{van Straaten} {et~al.}(2001){van Straaten}, {van der Klis},
  {Kuulkers}, \& {M{\'e}ndez}}]{vanSetal01}
{van Straaten}, S., {van der Klis}, M., {Kuulkers}, E., \& {M{\'e}ndez}, M.
  2001, \apj, 551, 907

\bibitem[{{Wallace} \& {Woosley}(1981)}]{WW81}
{Wallace}, R.~K. \& {Woosley}, S.~E. 1981, \apjs, 45, 389

\bibitem[{{Watts} \& {Strohmayer}(2006)}]{WS06}
{Watts}, A.~L. \& {Strohmayer}, T.~E. 2006, \mnras, 373, 769

\bibitem[{{Weinberg} {et~al.}(2006){Weinberg}, {Bildsten}, \& {Brown}}]{WBB06}
{Weinberg}, N.~N., {Bildsten}, L., \& {Brown}, E.~F. 2006, \apjl, 650, L119

\bibitem[{{Woosley} {et~al.}(2004){Woosley}, {Heger}, {Cumming}, {Hoffman},
  {Pruet}, {Rauscher}, {Fisker}, {Schatz}, {Brown}, \& {Wiescher}}]{Wetal04}
{Woosley}, S.~E., {Heger}, A., {Cumming}, A., {Hoffman}, R.~D., {Pruet}, J.,
  {Rauscher}, T., {Fisker}, J.~L., {Schatz}, H., {Brown}, B.~A., \& {Wiescher},
  M. 2004, \apjs, 151, 75

\bibitem[{{Woosley} \& {Taam}(1976)}]{WT76}
{Woosley}, S.~E. \& {Taam}, R.~E. 1976, \nat, 263, 101

\bibitem[{{Zingale} {et~al.}(2001){Zingale}, {Timmes}, {Fryxell}, {Lamb},
  {Olson}, {Calder}, {Dursi}, {Ricker}, {Rosner}, {MacNeice}, \&
  {Tufo}}]{Zetal01}
{Zingale}, M., {Timmes}, F.~X., {Fryxell}, B., {Lamb}, D.~Q., {Olson}, K.,
  {Calder}, A.~C., {Dursi}, L.~J., {Ricker}, P., {Rosner}, R., {MacNeice}, P.,
  \& {Tufo}, H.~M. 2001, \apjs, 133, 195

\end{thebibliography}

\end{document}